\documentclass[11pt]{article}
\usepackage{amsfonts,slashed}
\usepackage{arydshln}
\usepackage[greek,english]{babel}
\usepackage{cancel}
\usepackage{upgreek}
\usepackage{float}
\usepackage{url}
\usepackage{latexsym}
\usepackage{amsfonts}
\usepackage{amsmath}
\usepackage{epsfig}
\usepackage{latexsym,amssymb}
\usepackage{amsmath,amssymb,amsthm}
\usepackage{hyperref}
\usepackage{fontenc}
\usepackage{graphicx,tikz}
\usepackage{multirow}
\usepackage{xfrac}
\usepackage[T1]{fontenc}
\usepackage{textcomp}
\usepackage{lmodern}
\usepackage{stmaryrd}
\usepackage{cite}

\usepackage[margin=20pt,small]{caption}
\usepackage{subcaption}

\usepackage{ifpdf}
\ifx\pdfoutput\undefined
   \pdffalse
   \usepackage{cite}
 \else
   \pdfoutput=1
   \pdftrue
\fi
\DeclareFontFamily{OMS}{rsfs}{\skewchar\font'60}
\DeclareFontShape{OMS}{rsfs}{m}{n}{<-5>rsfs5 <5-7>rsfs7 <7->rsfs10 }{}
\DeclareSymbolFont{rsfs}{OMS}{rsfs}{m}{n}
\DeclareSymbolFontAlphabet{\Scr}{rsfs}

\numberwithin{equation}{section}
\def\be{\begin{equation}}
\def\ee{\end{equation}}
\def\ba{\begin{array}}
\def\ea{\end{array}}

\newcommand{\bea}{\begin{eqnarray}}
\newcommand{\eea}{\end{eqnarray}}

\def\={~=~}
\def\*{{}^*}

\def\P{\mathcal{P}}

\def\Tr{\mathrm{Tr}}

\newcommand\smallmath[2]{#1{\raisebox{\dimexpr \fontdimen 22 \textfont 2
      - \fontdimen 22 \scriptscriptfont 2 \relax}{$\scriptscriptstyle #2$}}}

\newcommand\smallotimes{\!\smallmath\mathbin\otimes\!}

\newcommand{\uA}{{\underline{A}}}
\newcommand{\uB}{{\underline{B}}}
\newcommand{\uC}{{\underline{C}}}
\newcommand{\uD}{{\underline{D}}}
\newcommand{\uE}{{\underline{E}}}
\newcommand{\uF}{{\underline{F}}}

\newcommand{\uM}{{\underline{M}}}
\newcommand{\uN}{{\underline{N}}}
\newcommand{\uP}{{\underline{P}}}

\newcommand{\MGrav}{\mathbb{M}_{\rm spin-2}}

\newcommand{\MScal}{\Big(\mathbb{M}_{\rm spin-0}\Big)}

\def\={~=~}
\def\*{{}^*}

\def\P{\mathcal{P}}

\def\Tr{\mathrm{Tr}}

\newcommand{\Sp}[1]{\mathrm{Sp}(#1)}
\newcommand{\SL}[1]{\mathrm{SL}(#1)}

\newcommand{\SO}[1]{\mathrm{SO}(#1)}
\newcommand{\En}[1]{\mathrm{E}_{#1(#1)}}

\newcommand{\fl}[1]{{\underline{#1}}}

\newcommand{\gL}{{\cal L}}
\newcommand{\cU}{{\cal U}}

\newcommand{\cY}{{\cal Y}}

\newcommand{\sV}{{\cal V}}
\newcommand{\cP}{{\cal P}}

\begin{document}
\begin{titlepage}
\vfill
\begin{flushright}
HU-EP-23/19
\end{flushright}

\vfill
\begin{center}
	{\LARGE \bf A holographic RG flow\\[1ex] 
	from the squashed to the round $S^7$
	}\\[1cm]
	
	{\large\bf Bastien Duboeuf\,$^{a}{\!}$
		\footnote{\tt bastien.duboeuf@ens-lyon.fr}, Michele Galli\,${}^{b}{\!}$
		\footnote{\tt gallimic@physik.hu-berlin.de}, Emanuel Malek\,${}^{b}{\!}$
		\footnote{\tt emanuel.malek@physik.hu-berlin.de}, Henning Samtleben\,${}^{a,c}{\!}$
		\footnote{\tt henning.samtleben@ens-lyon.fr} \vskip .8cm}
	
	{\it ${}^a$ ENSL, CNRS, Laboratoire de physique, F-69342 Lyon, France}\\ \ \\
	{\it  $^{b}$ Institut f\"{u}r Physik, Humboldt-Universit\"{a}t zu Berlin,\\
		IRIS Geb\"{a}ude, Zum Gro{\ss}en Windkanal 2, 12489 Berlin, Germany}\\ \ \\
	{\it  $^{c}$ Institut Universitaire de France (IUF)}\\ \ \\
	
\end{center}
\vfill

\begin{center}
	\textbf{Abstract}
	
\end{center}
\begin{quote}
We construct and analyze the domain wall solution in $D=11$ supergravity connecting the ${\cal N}=1$, AdS$_4\times S_{\rm squashed}^7$ vacuum to the ${\cal N}=8$,  AdS$_4\times S_{\rm round}^7$ vacuum. This domain wall describes the holographic renormalization group flow from an ${\rm Sp}(2)\times {\rm Sp}(1)$ symmetric UV fixed point to the ${\rm SO}(8)$ symmetric IR fixed point.
It breaks all supersymmetries which are (partially) restored at its endpoints. 
We show how recent techniques from exceptional field theory allow us to compute the quadratic couplings of all Kaluza-Klein fluctuations around the domain wall background, encoding all two-point correlators along the holographic RG flow.

\end{quote}
\vfill
\setcounter{footnote}{0}

\end{titlepage}

\newpage
\section{Introduction} \label{s:Intro}

The AdS/CFT correspondence is a powerful tool to analyse strongly-coupled CFTs. In this paper, we will focus on holographic RG flows. On the gravity side these correspond to domain-wall solutions with AdS vacua of string theory at the endpoint. The existence of such a solution between two different AdS vacua indicates that there is a renormalisation group flow between two conformal fixed points in the UV and IR, dual to the AdS vacua, something which may not even be directly visible in the QFT analysis. Moreover, the gravitational action encapsulates the QFT correlators along the flow, giving insight into strongly-coupled QFTs away from conformality, for example, into confining gauge theories \cite{Girardello:1999bd}.

Despite the promising technology, in practice, the analysis of holographic RG flows is limited to the use of lower-dimensional gauged supergravity models due to technical difficulties in working with the 10-/11-dimensional supergravity actions that capture the single-trace sector of the dual gauge theories. In particular, domain wall solutions corresponding to holographic RG flows are often constructed in such lower-dimensional gauged supergravities \cite{Freedman:1999gp,Bianchi:2001de,Berg:2001ty}, that can sometimes be uplifted via a consistent truncation to 10-/11-dimensional supergravity. However, in general, a consistent truncation may not exist, and even when it does, it can obscure important properties of the 10-/11-dimensional solutions. For example, the 10-/11-dimensional AdS solutions may have more supersymmetries than the corresponding solutions within the gauged supergravity. 

Even more restrictively, the holographic computation of correlation functions is limited to couplings between modes that are captured by the lower-dimensional gauged supergravity, ignoring the higher Kaluza-Klein modes in the supergravity. This is already true for 2-point functions. In fact, even the masses of higher Kaluza-Klein modes around AdS vacua could until recently not be studied, except for highly supersymmetric compactifications or those corresponding to coset spaces, rendering the computation of 2-point functions along an RG flow completely unfeasible.

The situation recently changed dramatically, thanks to the method of Kaluza-Klein spectroscopy \cite{Malek:2019eaz,Malek:2020yue} in Exceptional Field Theory (ExFT) \cite{Berman:2010is,Coimbra:2011ky,Coimbra:2012af,Berman:2011cg,Berman:2012vc,Hohm:2013pua}. This allows the computation of the full Kaluza-Klein spectrum around ``generalised parallelisable'' compactifications \cite{Lee:2014mla,Hohm:2014qga}. These are compactifications whose generalised tangent bundle in ExFT is trivial, i.e.\ they admit a global frame for the generalised tangent bundle. Examples of such compactifications include those that lead to consistent truncation of maximally gauged supergravities, but also their deformations beyond maximal gauged supergravity \cite{Duboeuf:2022mam}. As a result, the Kaluza-Klein spectrum around a variety of AdS vacua has been computed recently, including those with few or no symmetries, or indeed supersymmetries. This has led to various insights, such as infinitely many unprotected operators in strongly-coupled CFTs with finite conformal dimensions \cite{Bobev:2020lsk}, the compactness of conformal manifolds that appear non-compact in gauged supergravity computations \cite{Giambrone:2021zvp,Giambrone:2021wsm,Cesaro:2022mbu}, as well as instabilities triggered by higher Kaluza-Klein modes in non-supersymmetric AdS vacua \cite{Malek:2020mlk} that appear stable in gauged supergravity \cite{Fischbacher:2010ec} and the proof of perturbative stability of a number of AdS vacua in string theory \cite{Guarino:2020flh,Giambrone:2021wsm}.

In this paper, we will extend these techniques, which have so far only dealt with AdS vacua and their exactly marginal deformations, to compute the two-point supergravity couplings along an RG flow triggered by a relevant deformation. Crucially, this will be applicable to all Kaluza-Klein modes, not just those present within a consistent truncation. Moreover, we will show how this can be done for RG flows that do not even belong to a truncation to a maximal gauged supergravity.
Specifically, we focus on a holographic RG flow from the ${\cal N}=1$ AdS$_4 \times S^7_{\rm squashed}$ vacuum \cite{Awada:1982pk} to the ${\cal N}=8$ AdS$_4 \times S^7_{\rm round}$ vacuum of 11-dimensional supergravity. The squashing of the $S^7$ preserves only $\Sp{2} \times \Sp{1}$ isometries of the $\SO{8}$ isometries of the round $S^7$ and the corresponding 3-dimensional ${\cal N}=1$ CFT was discussed in \cite{Ooguri:2008dk}. Interestingly, while the round $S^7$ compactification admits a consistent truncation to 4-dimensional ${\cal N}=8$ gauged supergravity \cite{deWit:1984nz}, the ${\cal N}=1$ AdS$_4 \times S^7_{\rm squashed}$ vacuum does not sit within this consistent truncation. Moreover, the ${\cal N}=1$ Killing spinor does not belong to a subset of the ${\cal N}=8$ ones, but instead corresponds to higher Kaluza-Klein modes on the round $S^7$, in what is known as the space-invader scenario \cite{Duff:1986hr}.

Since the round and squashed vacua are not both part of ${\cal N}=8$ gauged supergravity, there has been some confusion in the literature about the potential existence of an RG flow between them. A first attempt stems from \cite{Ahn:1999dq}, where the relevant RG flow equations were written down, using an $\Sp{2} \times \Sp{1}$-invariant consistent truncation Ansatz \cite{Page:1984ad}. However, \cite{Campos:2000yu} argued that these equations do not admit a first-order formulation and, hence, that a flow is unlikely to exist. %
Later, consistent truncations on the squashed $S^7$ were constructed more systematically in \cite{Cassani:2011fu}, leading to a 4-dimensional ${\cal N}=4$ gauged supergravity. This can be further truncated to a ${\cal N}=1$ subsector and even further to a 2-scalar subsector, corresponding to that found in \cite{Page:1984ad}. Crucially, none of these truncations are subtruncations of the ${\cal N}=8$ gauged supergravity. Moreover, within these truncations, while the squashed $S^7$ vacuum is ${\cal N}=1$, the round $S^7$ appears as an ${\cal N}=0$ vacuum. This indicates that there is no supersymmetric $\Sp{2} \times \Sp{1}$-invariant flow in 11 dimensions, which can also be shown by an explicit computation.\footnote{We thank Nikolay Bobev for private communications on this point.}

This paper is laid out as follows. First, in section \ref{s:SquashedS7} we will review the squashed $S^7$ in 11-dimensional supergravity, as well as its embedding into the two-scalar truncation of\cite{Page:1984ad}. We numerically solve the RG flow equations and show the existence of a non-supersymmetric flow between the squashed and round $S^7$ vacua, clarifying earlier claims in the literature. In section \ref{s:Truncation}, we show how to embed the ${\cal N}=4$ \cite{Cassani:2011fu}, the ${\cal N}=1$ \cite{Cassani:2011fu}, and the two-scalar truncations of \cite{Page:1984ad} into ExFT. In particular, we will construct a generalised parallelisation for these consistent truncations, which, however, does not form an algebra under the generalised Lie derivative, reflecting the fact that the consistent truncations do not lead to ${\cal N}=8$ gauged supergravity. Nevertheless, the generalised parallelisation allows us to use the Kaluza-Klein spectroscopy to compute the two-couplings of the higher Kaluza-Klein modes along the RG flow from the squashed to round $S^7$, as we demonstrate in section \ref{s:KK}. Finally, we conclude with a discussion of our results and open questions in section \ref{s:Conclusions}.

\section{Squashed sphere, $S^7$ and domain wall in $D=11$ supergravity} \label{s:SquashedS7}

In this section, we review the round and the squashed $S^7$ vacua of $D=11$ supergravity and construct the interpolating domain wall solution. Both $S^7$ backgrounds are Freund-Rubin solutions of $D=11$ supergravity, preserving ${\cal N}=8$ and ${\cal N}=1$ supersymmetry, respectively~\cite{Awada:1982pk}. The general ansatz for the $D=11$ metric with an internal space $S^7$ preserving ${\rm Sp}(2)\times{\rm Sp}(1)$ isometries can be put into the form~\cite{Page:1983mke}
\begin{equation}
\begin{split}
ds^2 &= e^{-7u}\,ds^2_{(4)} + \frac14\,e^{2u} \,\left(
e^{3v}\,\Big(d\mu^2+\frac1{4}\,{\rm sin}^2\mu\,\sum_i \omega_i^2\Big)
+\frac14\,e^{4v}\,\sum_i \left(\nu_i+{\rm cos}\,\mu\,\omega_i\right)^2\right)
\,,
\end{split}
\label{eq:domain_wall}
\end{equation}
with $S^7$ size parameter $u$ and squashing parameter $v$ that are taken as scalar functions
over the four-dimensional space-time. The one forms $\omega_i=\sigma_i-\Sigma_i$, $\nu_i=\sigma_i+\Sigma_i$, satisfy
\begin{equation}
d\sigma_i= -\frac12\,\varepsilon_{ijk}\,\sigma_j\wedge \sigma_k\,,\quad
d\Sigma_i= -\frac12\,\varepsilon_{ijk}\,\Sigma_j\wedge \Sigma_k
\,.
\end{equation}
The 4-form flux for these solutions is of the form 
\begin{equation}
F_{\mu\nu\rho\sigma} = Q\,e^{-21u}\,\varepsilon_{\mu\nu\rho\sigma}
\,,
\label{eq:F4}
\end{equation}
with conserved charge $Q$. The ansatz \eqref{eq:domain_wall}, \eqref{eq:F4} is, in fact, a consistent truncation of $D=11$ supergravity as shown in~\cite{Page:1983mke}. More precisely, plugging \eqref{eq:domain_wall}, \eqref{eq:F4}, into the $D=11$ field equations leads to field equations that are obtained from the $D=4$ Lagrangian 
\begin{equation}\label{eq:potential}
\begin{split}
|g|^{-1/2}\,{\cal L}_{(4)}^{0} &=
R_{(4)}
-\frac{63}{2}\,\partial_\mu u \,\partial^\mu u -21\,\partial_\mu v \,\partial^\mu v - V_{\rm pot}\,, \\
V_{\rm pot} &= -6\,e^{-9u+4v}-48\,e^{-9u-3v}+12\,e^{-9u-10v}+2\,Q^2\,e^{-21u}
\,.
\end{split}
\end{equation}
The parameter $Q$ may be absorbed into a shift of $u$ together with a rescaling of the $D=4$ metric $g_{\mu\nu}$. In the following, we set $Q=3$\,.

Extremisation of the potential $V_{\rm pot}$ from \eqref{eq:potential} yields two critical points 
corresponding to ${\rm AdS}_4\times S^7$ solutions of $D=11$ supergravity, 
with the round and the squashed sphere located at
\begin{equation}
\begin{split}
\mbox{round } S^7 &: \quad u=0\;,\;\;v=0\;,\quad\ell_{\rm round} = \tfrac12 \,,\\
\mbox{squashed } S^7 &: \quad u=u_0\equiv\tfrac{5}{42} {\rm ln}\,5 -\tfrac16{\rm ln}\,3 \;,\;\;v=v_0\equiv\tfrac17 {\rm ln}\,5 \,,\quad\ell_{\rm squashed} = \tfrac{5^{5/4}}{2\cdot 3^{7/4}} \,,
\end{split}
\label{eq:S7S7}
\end{equation}
and the respective AdS lengths given by $\ell=\sqrt{-\frac{6}{V_{\rm pot}}}$. With the standard ansatz for an interpolating domain wall solution
\begin{equation}
u=u(r)\,,\;\;v=v(r)\,,\quad
ds_{(4)}^2 = dr^2 + e^{2A(r)}\, \eta_{ij}\,dx^i dx^j \,,\quad
 i, j=1, 2, 3\,,
\end{equation}
variation of \eqref{eq:potential} yields the equations
\begin{equation}
\begin{split}
u'' +3\,A'u' &= -6 \,e^{-21 u}-\frac{12}{7} \,e^{-9 u-10 v}+\frac{48}{7} \,e^{-9 u-3 v}+\frac{6}{7} \,e^{-9 u+4 v}\,, \\
v'' +3\,A'v'&= -\frac{20}{7} \,e^{-9 u-10 v}+\frac{24}{7} \,e^{-9 u-3 v}-\frac{4}{7} \,e^{-9 u+4 v} \,,\\
3\,(A')^2-\frac{63}{4}\,(u')^2-\frac{21}{2}\,(v')^2 &= -9 \, e^{-21 u}-6 \,e^{-9 u-10 v}+24 \,e^{-9 u-3 v}+3 \,e^{-9 u+4 v} \,.
\end{split}
\label{eq:flow-equations}
\end{equation}
The existence of a domain wall solution to these equations, interpolating between the two vacua \eqref{eq:S7S7}, was discussed in \cite{Ahn:1999dq}, and later questioned in \cite{Campos:2000yu}.  Its holographic interpretation was further elaborated in \cite{Ooguri:2008dk}. The domain wall represents a holographic renormalisation group flow from an ${\cal N}=1$ superconformal UV fix point (dual to the squashed $S^7$) to an ${\cal N}=8$, ${\rm SO}(8)$ symmetric IR fix point (dual to the round $S^7$).
Unlike most explicitly known domain wall solutions, there is no description of the flow equations connecting \eqref{eq:S7S7} in terms of first order differential equations and a superpotential. More precisely, \cite{Campos:2000yu} noted that the potential \eqref{eq:potential} can be written in terms of a superpotential $W$ as
\begin{equation}
\begin{split}
V_{\rm pot} &=\frac{16}{63}\,(\partial_u W)^2+\frac{8}{21}\,(\partial_v W)^2-12\,W^2 \,,\\
W &= -\frac3{\sqrt{8}}\,e^{-9u/2}\left( e^{2v}+2\,e^{-5v}-e^{-6u}\right) \,.
\label{eq:superpotential}
\end{split}
\end{equation}
However only the squashed $S^7$ represents a critical point of this superpotential. With hindsight, this is a manifestation of the fact that the Lagrangian \eqref{eq:potential} lives within an ${\cal N}=1$ four-scalar truncation of $D=11$ supergravity~\cite{Cassani:2011fu}, in which the round sphere appears as an ${\cal N}=0$ vacuum, since all the ${\cal N}=8$ massless gravitinos around this vacuum are truncated out.\footnote{In the notation of \cite{Cassani:2011fu}, all this is on the $k<0$ branch.} Accordingly, the round $S^7$ does not correspond to a critical point of the associated superpotential \eqref{eq:superpotential}.

\begin{figure}[tb]
\center
\includegraphics[scale=0.3]{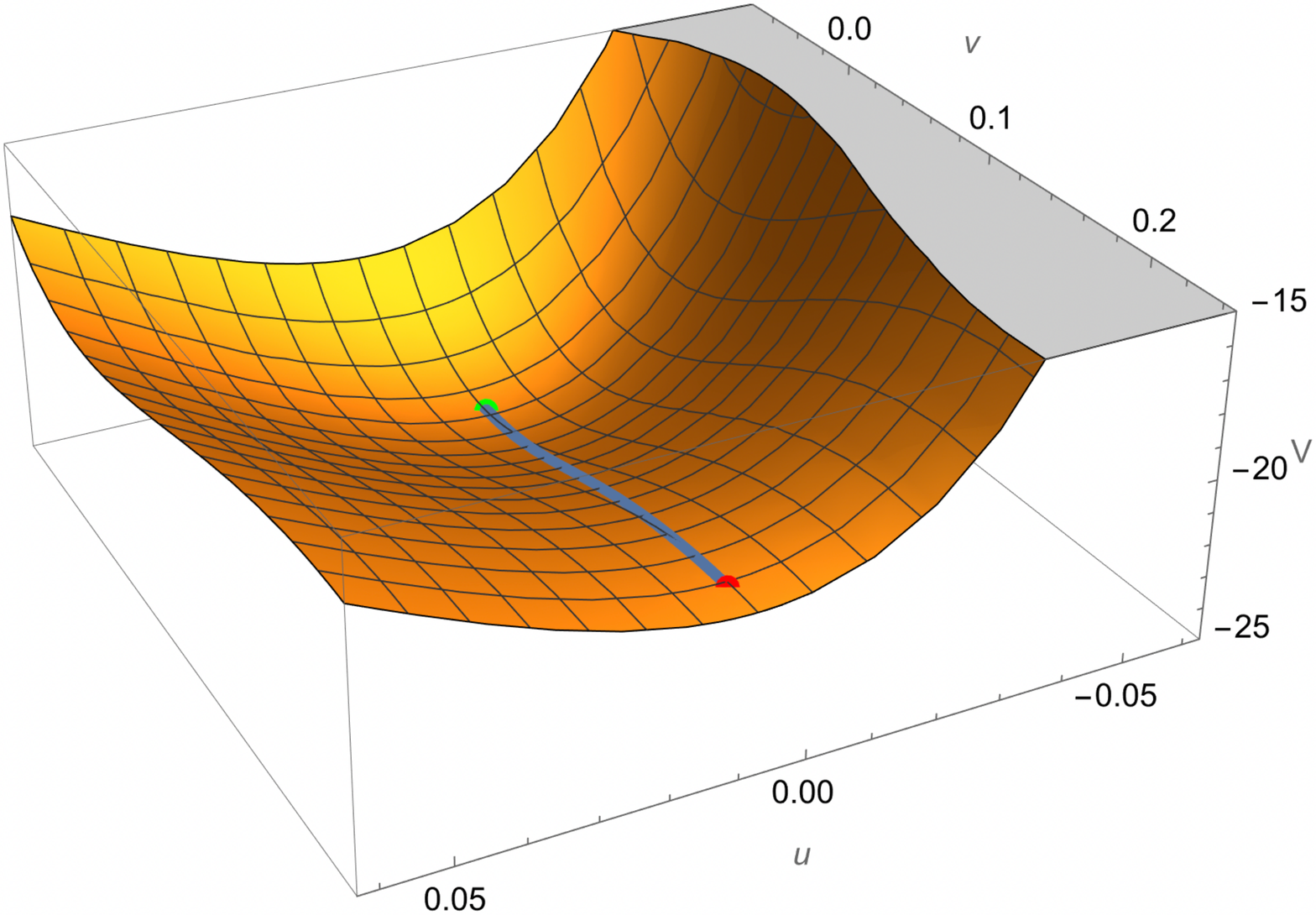}
\caption{The domain wall (blue line) in the scalar potential of~\eqref{eq:potential}. 
The green and the red dots represent the round and the squashed sphere \eqref{eq:S7S7}, respectively.}
\label{fig:flow}
\end{figure}

The interpolating domain wall solution can be found numerically, by solving equations \eqref{eq:flow-equations} and fine-tuning the initial conditions. To this end, it is instructive to first study the general asymptotic behaviour of the scalar fields. As usual, this is correlated with the conformal dimensions of the associated dual operators, given by
\begin{equation} \label{eq:DeltasUVIR}
\begin{split}
{\cal O}_u&: \quad \Delta_{\rm UV}=6= \Delta_{\rm IR}
\,,\\
{\cal O}_v&: \quad \Delta_{\rm UV}=\tfrac53\,,\;\;  \Delta_{\rm IR}=4\,,
\end{split}
\end{equation}
for the scalar fields. Introducing the radial coordinate
\begin{equation}
\rho= e^{-2r/{\ell_{\rm squashed}}}
\;,
\end{equation}
and expanding the flow equations \eqref{eq:flow-equations} near the UV boundary $\rho=0$, we find the following asymptotic expansions of its general regular solution
\begin{equation}\label{eq:Asymptotics}
\begin{split}
& u(\rho) = \tfrac16\,{\rm ln}\,\tfrac{5^{5/7}}{3}+
\rho^{\frac86} \,\Omega_{1,8} + \rho^{\frac96} \,\Omega_{1,9} + \rho^{\frac{10}{6}} \, \Omega_{1,10} + \dots + \rho^{18/3}\,\underline{\Omega}_{1,18} + \dots \;, \\
& v(\rho) = \tfrac17\,{\rm ln}\,5+
\rho^{\frac46} \,\underline{\Omega}_{2,4} + \rho^{\frac56} \,\underline{\Omega}_{2,5} + \rho^{\frac{8}{6}} \,\Omega_{2,8} + \rho^{\frac{9}{6}} \, \Omega_{2,9} + \rho^{\frac{10}{6}} \, \Omega_{2,10} + \dots\;, \\
& 2A(\rho) = - \log\rho + A_8 \,\rho^\frac86 +  A_9\, \rho^\frac96 + A_{10}\, \rho^\frac{10}{6} + \dots \;.
\end{split}
\end{equation}
Equations (\ref{eq:flow-equations}) fix all coefficients $\Omega_{i,j}$ in the above expressions,
except $\underline{\Omega}_{2,4}$, $\underline{\Omega}_{2,5}$, and $\underline{\Omega}_{1,18}$ which source the others. 
Explicitly, we find that the coefficients of the lowest powers are given by
\begin{equation}
\begin{split}
\Omega_{1, 8} &= - \tfrac{6}{17} \,\underline{\Omega}_{2,4}^2 \,,\quad\;
 \Omega_{1, 9} =- \tfrac{20}{27} \,\underline{\Omega}_{2,4} \,\underline{\Omega}_{2,5}\,,
\\
\Omega_{2,8} &=- \tfrac{15}{4} \underline{\Omega}_{2,4}^2  \,,\quad\;\,
 \Omega_{2, 9}  = -\tfrac{9}{2}\,\underline{\Omega}_{2,4} \,\underline{\Omega}_{2,5}\,,
\\
	A_8 &=- \tfrac{21}{4} \,\underline{\Omega}_{2,4}^2\,,\quad\;\;\;
	A_9 =- \tfrac{280}{27} \,\underline{\Omega}_{2,4}\,\underline{\Omega}_{2,5}\,.
\end{split}
\end{equation}
In \eqref{eq:Asymptotics}, we have imposed regularity at $\rho=0$, which sets to zero a potential $\rho^{-\frac32}$ term in the $u$ expansion, allowed by equations \eqref{eq:flow-equations}. The general solution of \eqref{eq:flow-equations}, regular at $\rho=0$, thus carries three independent constants. It furthermore admits the scaling symmetry $\rho\rightarrow\lambda\rho$, which can be used to set $\underline{\Omega}_{2,4}=-1$\,. For the interpolating solution, the remaining two coefficients are then fixed by further demanding regularity at the other end of the flow $\rho\rightarrow\infty$.

\begin{figure}[bt]
\center
\includegraphics[scale=0.27]{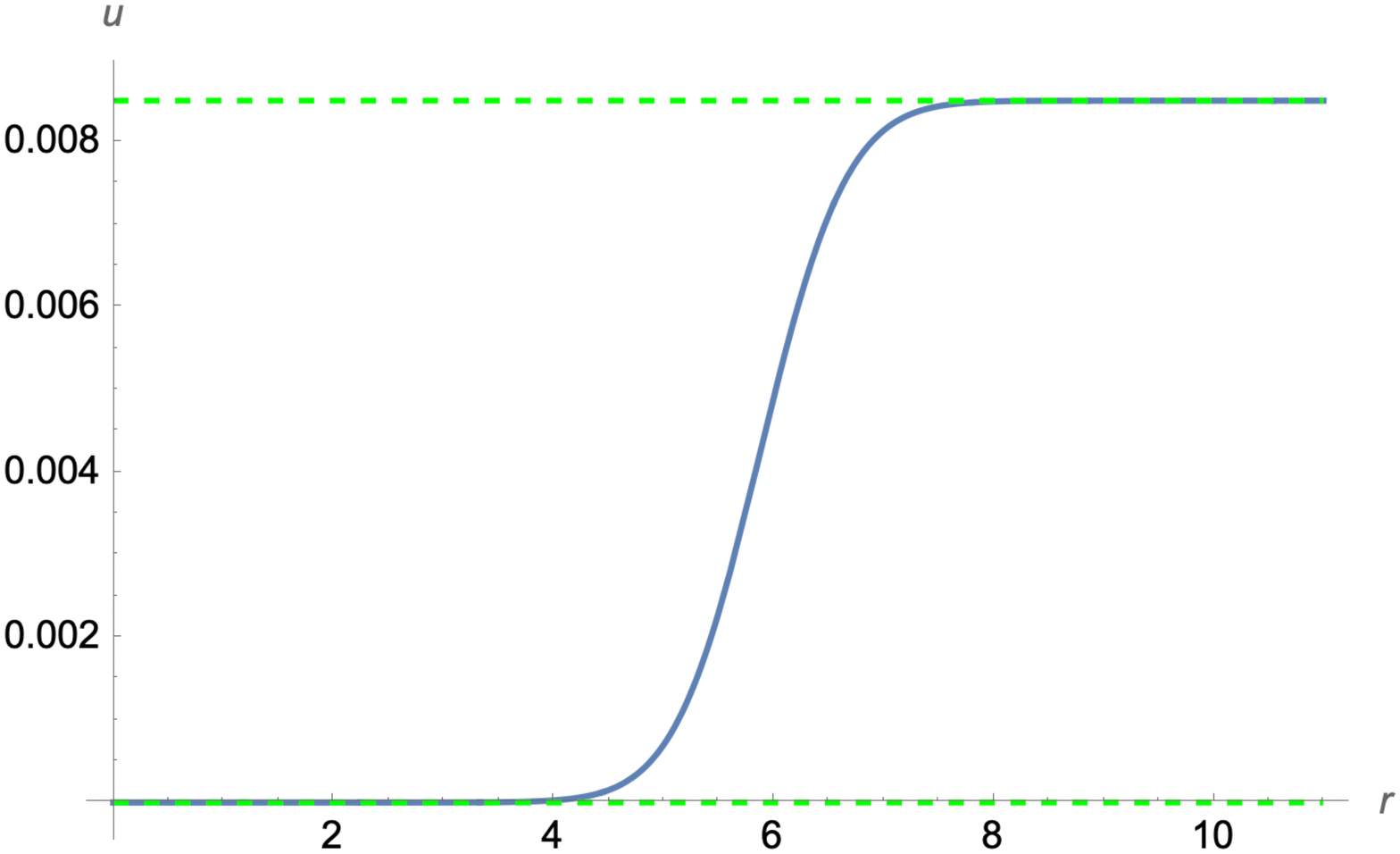}
\includegraphics[scale=0.27]{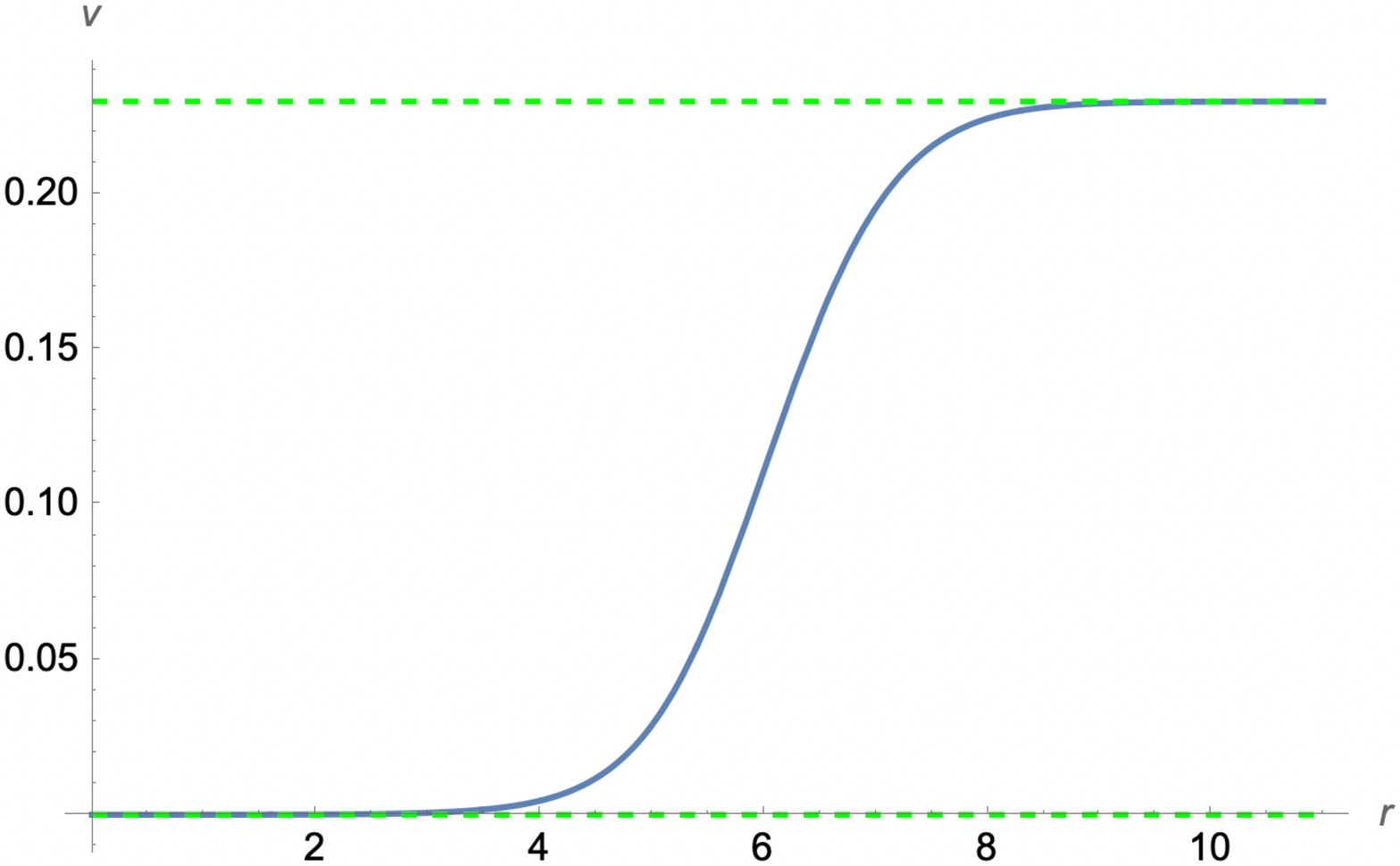}
\includegraphics[scale=0.27]{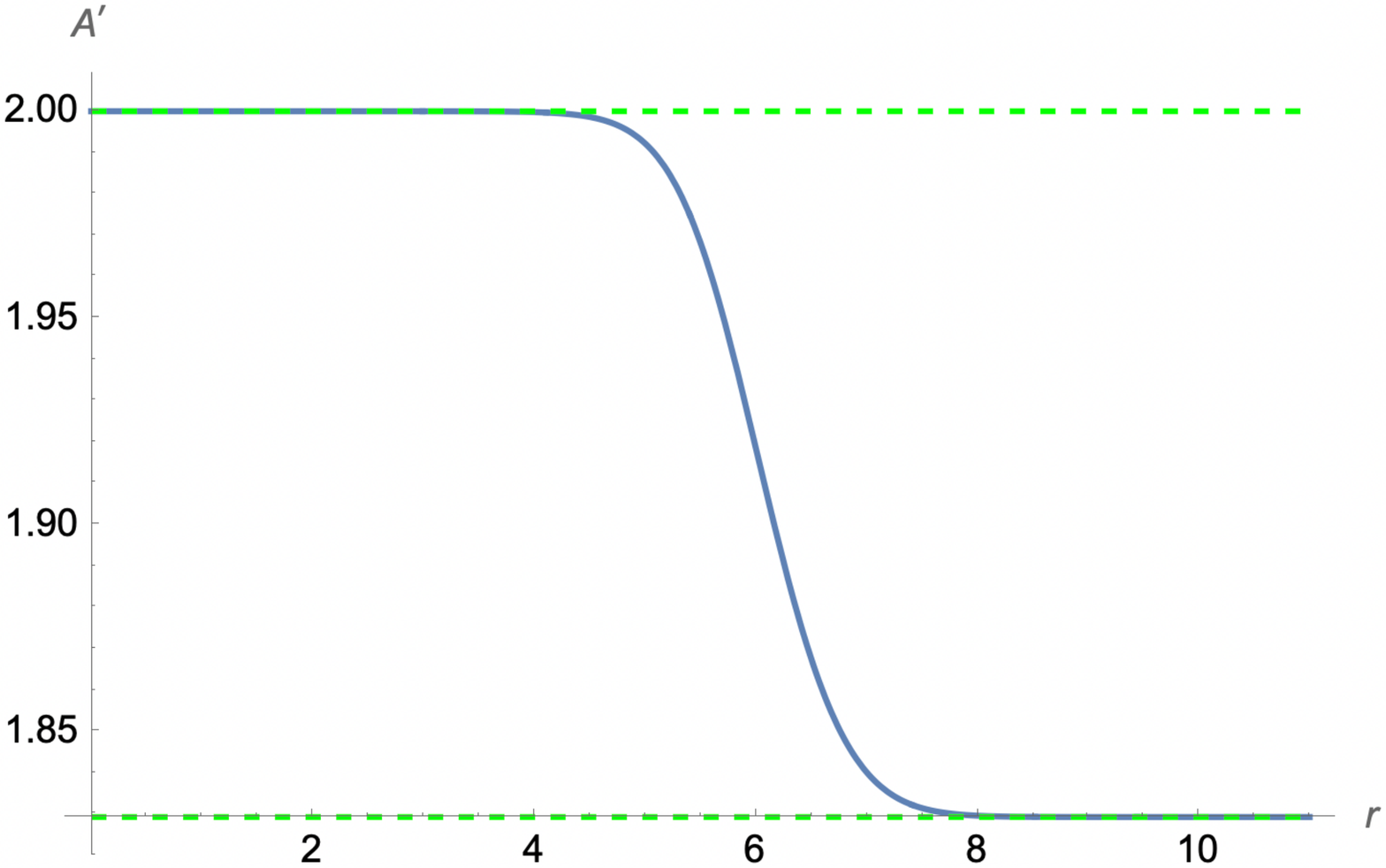}
\caption{Numerical solutions for the two scalar fields, as well as the derivative $A'$. The horizontal green dashed line represent the asymptotic values of the various fields. The UV boundary is located at $r\rightarrow\infty$, the IR boundary is at $r\rightarrow-\infty$.}
\label{fig:scalars}
\end{figure}

\begin{figure}[tb]
\center
\includegraphics[scale=.54]{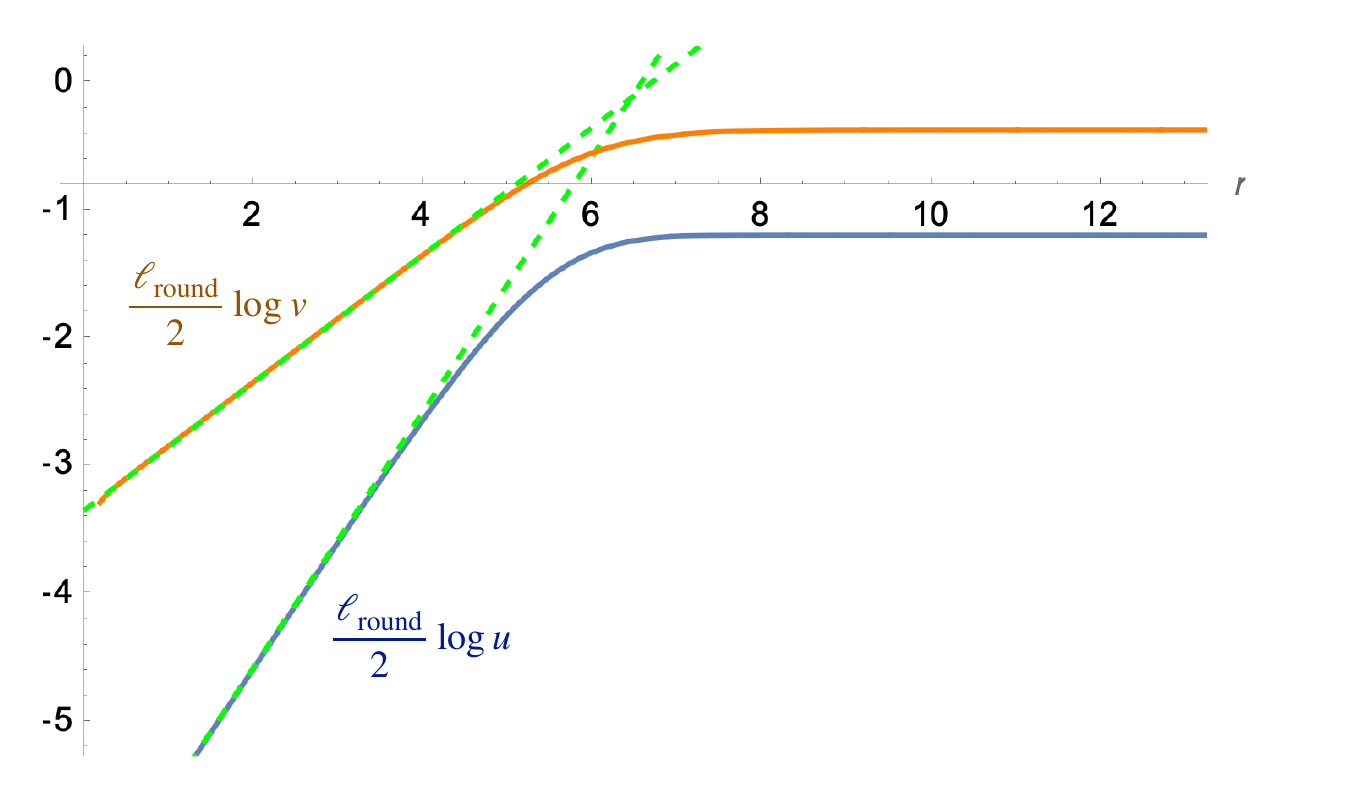}
\includegraphics[scale=.50]{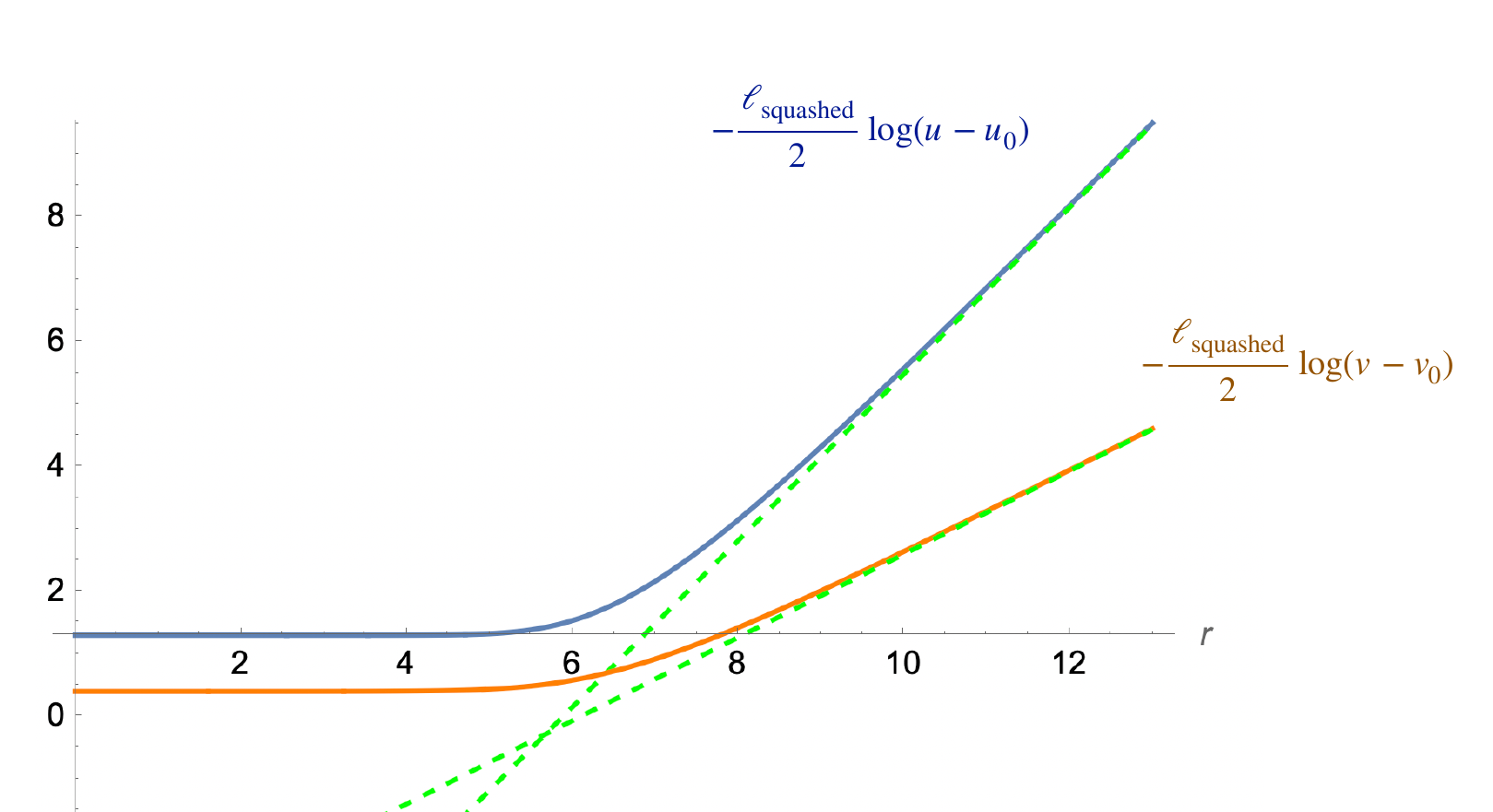}
\caption{Asymptotic behavior of the numerical solution. In the UV ($r\rightarrow\infty$), the green dashed line have respectively a slope of $\frac43$ for $u$ and $\frac23$ for $v$. This matches the asymptotic expansion  \eqref{eq:Asymptotics}. In the IR ($r\rightarrow-\infty$), the green dashed line have respectively a slope of $1$ for $u$ and $\frac12$ for $v$. This matches the asymptotic expansion  \eqref{eq:AsymptoticsIR}.}
\label{fig:Asymptotics}
\end{figure}

Imposing regularity at both ends of the flow, we have solved equations \eqref{eq:flow-equations} numerically, and plot the result in Figures~\ref{fig:flow},~\ref{fig:scalars}. As expected, the solution is of kink type for the scalar fields $u$, $v$, as well as for the derivative $\partial_r A$\,. In particular, with $\underline{\Omega}_{2,4}=-1$, we find for the coefficient $\underline{\Omega}_{2,5}$ the approximate numerical value
\begin{equation}
\underline{\Omega}_{2,5} \approx -1.4
\;.
\end{equation}
In Figure~\ref{fig:Asymptotics}, we plot the asymptotics of the scalar fields, which confirms the UV expansion \eqref{eq:Asymptotics}, and in particular the fact that the leading coefficient $\underline{\Omega}_{2,4}$ is non-vanishing. This is the expected power-law behaviour
\begin{equation}
v(\rho)-v_0 \propto \rho^{\frac{3-\Delta}{2}}\,,
\end{equation}
consistent with the interpretation that the holographic dual of this domain wall solution is an operator deformation (rather than a vev) of the UV CFT \cite{Klebanov:1999tb,Bianchi:2001de}, by a relevant operator of conformal dimension $\Delta_{\rm UV}=\frac53$\,.

Similarly, one can work out the asymptotic behaviour in the IR. With the new radial variable $\tilde\rho= e^{2r/{\ell_{\rm round}}}$, the expansion of a general solution of \eqref{eq:flow-equations}, regular at the IR boundary $\tilde\rho=0$, is given by
\begin{equation}\label{eq:AsymptoticsIR}
\begin{split}
& u(\tilde\rho) =
{\omega}_{1,2}\,\tilde\rho + \underline\omega_{1,3}\,\tilde\rho^{3/2}+ \psi_{1,3}\,\tilde\rho^{3/2}\,{\rm log}\,\rho
+ \omega_{1,4}\,\tilde\rho^{2}
+\dots
\;, \\
& v(\tilde\rho) = \underline{\omega}_{2,1}\,\tilde\rho^{1/2} + \omega_{2,2}\,\tilde\rho
+\omega_{2,3}\,\tilde\rho^{3/2}+\omega_{2,4}\,\tilde\rho^2 + \psi_{2,4}\,\tilde\rho^{2}\,{\rm log}\,\rho 
 + \dots\;, \\
& 2A(\tilde\rho) =  \log\tilde\rho + \tilde A_2 \,\tilde\rho + \tilde A_3 \,\tilde\rho^{3/2} +  \tilde A_4 \,\tilde\rho^2 + \dots \;,
\end{split}
\end{equation}
in accordance with the IR conformal dimensions \eqref{eq:DeltasUVIR}. Here, imposing regularity at $\tilde\rho=0$ has set to zero two of the free coefficients of the general solution of \eqref{eq:flow-equations}. All other coefficients in the expansion \eqref{eq:AsymptoticsIR} are then determined in terms of the remaining two free coefficients $\underline{\omega}_{1,3}$, $ \underline{\omega}_{2,1}$, e.g.
\begin{equation}
\begin{split}
\omega_{1, 2} &=\tfrac32\,\underline{\omega}_{2,1}^2 \,,\quad\quad\,
\psi_{1,3}=11\,\underline{\omega}_{2,1}^3  \,,\quad\;\;\,
\omega_{1, 4} =-\tfrac{15525}{56}\,\underline{\omega}_{2,1}^4\,,\quad
\\
\omega_{2,2}&=-\tfrac{11}{2}\,\underline{\omega}_{2,1}^2 \,,\quad \omega_{2, 3}  = \tfrac{5261}{168}\,\underline{\omega}_{2,1}^3 \,,\quad
\omega_{2,4}=-\tfrac32\,\underline{\omega}_{2,1}\underline{\omega}_{1,3}-\tfrac{10583}{56}\,\underline{\omega}_{2,1}^4 \,,
\quad
\psi_{2,4}=-\tfrac{33}{2}\,\underline{\omega}_{2,1}^4 \,,
\\
\tilde A_2 &= -\tfrac{21}{4}\,\underline{\omega}_{2,1}^2\,,\quad\,\;  \tilde A_3 =\tfrac{154}{3}\,\underline{\omega}_{2,1}^3\,,\;\;\,\,
\quad  \tilde A_4 =-\tfrac{27081}{64}\,\underline{\omega}_{2,1}^4\,.
\end{split}
\end{equation}
Again, the leading terms are confirmed by the plots of the numerical solution in Figure~\ref{fig:Asymptotics}. 
Regularity at the UV end of the flow finally fixes $\underline{\omega}_{1,3}$ as a function of $ \underline{\omega}_{2,1}$. From the above numerical domain wall solution,
we find the approximate value
\begin{equation}
\frac{\underline{\omega}_{1,3}}{(\underline{\omega}_{2,1})^{3}} ~\approx~ -239 
\;,
\end{equation}
for the combination invariant under the scaling symmetry $\tilde\rho\rightarrow\lambda\tilde\rho$\,.

\section{Generalised parallelisation of the domain wall in ExFT} \label{s:Truncation}

In this section, we will identify the AdS$_4\times S_{\rm squashed}^7$ background and the domain wall \eqref{eq:domain_wall} within ExFT, i.e.\ within the duality-covariant formulation of $D=11$ supergravity. This allows us to construct consistent truncations around this vacuum as well as to compute the quadratic couplings of Kaluza-Klein fluctuations around the domain wall background in section \ref{s:KK}.

The relevant ExFT is based on the exceptional group E$_{7(7)}$ and has been constructed in \cite{Hohm:2013uia} to which we refer for details. Its structures relevant to our construction have been reviewed in~\cite{Malek:2020yue}. From a most pragmatic point of view, E$_{7(7)}$ ExFT can be viewed as a reformulation of $D=11$ supergravity (upon splitting the coordinates into 11=4+7) in terms of variables that mimic the field content of $D=4$ maximal supergravity. In particular, the internal components of the $D=11$ metric, three-form, and six-form are assembled into an ${\rm E}_{7(7)}$ group matrix, the generalized vielbein, defined as
\begin{equation} \label{V56}
	\begin{split}
		{\cal V} &\equiv
		{\rm exp}\left[A_{klmnpq}\, t_{(+4)}^{klmnpq} \right]
		{\rm exp}\left[A_{kmn}\,t_{(+2)}^{kmn}\right]
		\,V_{{\rm GL}(7)} \,,
	\end{split}
\end{equation}
i.e.\ as a coset representative of ${\rm E}_{7(7)}/{\rm SU}(8)$ in a particular triangular gauge. Here, $V_{{\rm GL}(7)} \in {\rm GL}(7)\subset {\rm E}_{7(7)}$ is the internal block of the 11D vielbein (up to some power of its determinant), 
while $A_{kmn}$ and $A_{klmnpq}$ denote the internal components of the $D=11$ three-form and six-form, respectively, with $k, l, m = 1, \ldots, 7$.
The $t_{(+n)}$ refer to the E$_{7(7)}$ generators of positive grading in the algebra decomposition
\begin{eqnarray}
\mathfrak{e}_{7(7)} &{\longrightarrow}& 7'_{-4}  \oplus 35_{-2} \oplus \mathfrak{gl}(7)_0 \oplus 35'_{+2} \oplus  7_{+4} \,.
\end{eqnarray}
All generators in \eqref{V56} are evaluated in the fundamental ${\bf 56}$ representation of ${\rm E}_{7(7)}$. For a 
Freund-Rubin (FR) background (i.e.\ a solution with $A_{kmn}=0$), the parametrization (\ref{V56}) only involves generators within the $\mathfrak{sl}(8)$ subalgebra of $\mathfrak{e}_{7(7)}$ and the associated generalized vielbein can be represented as an $8\times8$ matrix
\begin{equation}
{\cal V}_{\rm FR} \in {\rm SL}(8)\big/ {\rm SO}(8) \,.
\label{eq:Freund-Rubin}
\end{equation}

For example, for the round $S^7$ with 7-form flux, the generalized vielbein \eqref{V56} takes the form
\begin{equation}
{\cal V}_{S^7} = {\rm exp}\left[-6\, t_{(+4)}^{lmnpqr}\,\zeta^k \mathring{\omega}_{klmnpqr} \right]
\begin{pmatrix} \mathring{\omega}^{3/4} & 0\\ 0 & \mathring{\omega}^{-1/4} \, \mathring{e}_m{}^i \end{pmatrix} 
\in {\rm SL}(8)
\,,
\label{eq:VS7}
\end{equation}
where $\mathring{e}_m{}^i$ is the $S^7$ vielbein, $\mathring{\omega}_{klmnpqr}$ is the associated volume form, $\mathring{\omega} \equiv {\rm det} \,\mathring{e}_m{}^i$, $\zeta^k$ is a vector field with $\mathring\nabla_k\zeta^k =1$ and $t_{(+4)}$ is evaluated in \eqref{eq:VS7} in the $\mathbf{8}$ representation of $\SL{8}$.

\subsection{Generalised Leibniz parallelisation and Kaluza-Klein fluctuations} \label{s:GLPandMass}
The round $S^7$ with 7-form flux \eqref{eq:VS7} has been extensively studied in the ExFT framework due to its interesting properties. In particular, it is generalised Leibniz parallelisable, i.e.\ the generalized vielbein \eqref{eq:VS7} is related by an ${\rm SO}(8)$ gauge transformation
\begin{equation}
{\cal V}_{S^7} = \mathring{U} \,S_{{\rm SO}(8)}
\;,
\label{eq:VUS7}
\end{equation}
to a twist matrix $\mathring{U}$, such that the associated generalized frame field ${\cal U}=\rho^{-1} \mathring{U}^{-1}$
(here, $\rho=\mathring\omega^{-1/2}$) evaluated in the $\mathbf{56}$ representation, satisfies the condition
\begin{equation} \label{eq:GenLeib}
	\gL_{\cU_{{\fl{A}}}} \cU_{\fl{B}}{}^M = X_{\fl{AB}}{}^{\fl{C}}\, \cU_{\fl{C}}{}^M \,,
\end{equation}
where $X_{\fl{AB}}{}^{\fl{C}}$ is the constant \emph{intrinsic torsion}. The generalized diffeomorphisms of the $\En{7}$ ExFT are defined as \cite{Coimbra:2011ky,Berman:2012vc}
\begin{equation} \label{eq:gen_diff_M}
	\gL_\Lambda V^M = \Lambda^N \partial_N V^M - 12\, \partial_K \Lambda^L \, (t_\alpha)^K{}_L (t^\alpha)^{M}{}_N \, V^N + \tfrac12 V^M \partial_N \Lambda^N \,.
\end{equation}
Explicitly, the twist matrix $\mathring{U}$ for the round $S^7$ is given by \cite{Lee:2014mla,Hohm:2014qga}
\begin{equation}
	\mathring{U}_{\underline{m}}{}^{a}({\cal Y}) =
	\begin{pmatrix}
	\mathring\omega^{3/4}\left({\cal Y}^a 
-6\,\zeta^n \partial_n {\cal Y}^a \right) \\
	 \mathring\omega^{-1/4}\,\partial_m {\cal Y}^a
	\end{pmatrix}  \in {\rm SL}(8)
	\,,\qquad
	\underline{m}=\{0,m\}\,, \quad a = \{1, \ldots, 8\} \,,
	\label{eq:Uround_alpha}
\end{equation}
in terms of the fundamental sphere harmonics ${\cal Y}^a{\cal Y}^a=1$.

As a consequence of \eqref{eq:GenLeib}, $D=11$ supergravity admits a maximally supersymmetric consistent truncation around the round $S^7$ to $D=4$, ${\cal N}=8$, ${\rm SO}(8)$ gauged supergravity \cite{deWit:1982bul,deWit:1986iy}. In particular, the embedding of the 70 scalar fields of $D=4$, ${\cal N}=8$ supergravity into the eleven-dimensional theory is compactly described in terms of (\ref{V56}) as
\begin{equation}
{\cal V}(x,y) = \mathring{U}(y) \,V(x) \,,
\label{eq:VUV}
\end{equation}
with the twist matrix $\mathring{U}$ from (\ref{eq:Uround_alpha}), and the ${\rm E}_{7(7)}/{\rm SU}(8)$ valued matrix
$V(x)$ carrying the 70 scalar fields of the $D=4$ theory.

Moreover, the existence of a globally defined generalized frame allow us to recover the full Kaluza-Klein spectrum around this background by virtue of the universal formulae derived in \cite{Malek:2019eaz,Malek:2020yue,Duboeuf:2022mam}. Here, we just recall the result. While (\ref{eq:VUV}) describes the embedding of the lowest Kaluza-Klein multiplet into $D=11$, the embedding of the infinite Kaluza-Klein tower of scalar fluctuations $j^{I,\Sigma}(x)$ is most compactly described as
\begin{equation} 
{\cal V}(x,y) = \mathring{U}(y)\,\Big( \mathbb{I}+ \cP_{I} {\cal J}^I(x,y) \Big)
 = \mathring{U}(y)\,\Big( \mathbb{I}+ \cP_{I} \sum_\Sigma  j^{I,\Sigma}(x) \,  \cY_\Sigma(y) \Big)
\;.
\label{eq:scalarFluc}
\end{equation}
Here, $\cY_\Sigma(y)$ label the scalar harmonics on $S^7$, and ${\cal P}_I$, $I = 1, \ldots, 70$, are the 70 non-compact generators of ${\rm E}_{7(7)}$.
After subtraction of the non-physical Goldstone modes, the action of the mass operator on the scalar fluctuations ${\cal J}^I(x,y)$ is given by
\begin{equation} \label{eq:MassMatrixScalarClean}
	\begin{split}
		\MScal{}^I{}_J &= \mathbb{M}_{(0)}{}^{I}{}_{J} + \left( \mathbb{N}^{I}{}_{J}{}^{\fl{C}} - \mathbb{N}_J{}^I{}^{\fl{C}} \right) \partial_{\fl{C}} + \partial_{\fl{C}} \mathbb{N}^{I}{}_{J}{}^{\fl{C}} + \delta^{I}_{J}\, \MGrav \,,
	\end{split}
\end{equation}
with
\begin{equation}
	\begin{split}
		\mathbb{M}_{(0)}{}^{I}{}_{J} &= X_{\uA\uE}{}^{\uF} X_{\uB\uF}{}^{\uE} \, (\cP^I \cP_J)_{\uA}{}^{\uB} \\
		& \quad + \tfrac17 \left( X_{\uA\uE}{}^{\uF} X_{\uB\uE}{}^{\uF} + X_{\uE\uA}{}^{\uF} X_{\uE\uB}{}^{\uF} + X_{\uE\uF}{}^{\uA} X_{\uE\uF}{}^{\uB}  \right) (\P^I \cP_J)_{\uA}{}^{\uB} \\
		& \quad + \tfrac27 \left( X_{\uA\uC}{}^{\uE} X_{\uB\uD}{}^{\uE} - X_{\uA\uE}{}^{\uC} X_{\uB\uE}{}^{\uD} - X_{\uE\uA}{}^{\uC} X_{\uE\uB}{}^{\uD} \right) (\cP^I)_{\uA}{}^{\uB}\, (\cP_J)_{\uC}{}^{\uD} \\
		& \quad + \tfrac16 \, X_{\fl{A}}{}^{I}\, X_{\fl{A},J} \,,
\\[1ex]
		\mathbb{N}^I{}_J{}^{\fl{C}} &= - 2 X_{\fl{A}}{}^I \cP_{J,\fl{C}}{}^{\fl{A}}- 2 X_{\fl{A}J} \cP^I{}_{\fl{C}}{}^{\fl{A}} - [ \cP^I,\, \cP_J ]_{\fl{A}}{}^{\fl{B}} \left( X_{\fl{C}\fl{B}}{}^{\fl{A}} + 6\, X_{\fl{A}\fl{B}}{}^{\fl{C}} \right) \,,
\\[1ex]
\MGrav &= - \partial_{\fl{A}} \partial_{\fl{A}} \,.
	\end{split}
	\label{eq:MNM}
\end{equation}

Evaluating this formula with the intrinsic torsion $X_{\uA\uB}{}^{\uC}$ from \eqref{eq:GenLeib} and an explicit basis for $\cY_\Sigma(y)$ allows us to derive the Kaluza-Klein spectrum for the full infinite tower of scalar fluctuations.	Specifically, for the round $S^7$ vacuum, this reproduces the result for the scalar Kaluza-Klein towers~\cite{Sezgin:1983ik,Biran:1983iy,Casher:1984ym}
\begin{equation}
\sum_\ell \Big(
[\ell+2,0,0,0]_{\frac\ell2+1} + 
[\ell-2,0,0,0]_{\frac\ell2+5} + 
[\ell-2,2,0,0]_{\frac\ell2+3}  + 
[\ell,0,2,0]_{\frac\ell2+2}  + 
[\ell-2,0,0,2]_{\frac\ell2+4}  
\Big)
\,,
\label{eq:towersS7}
\end{equation}
in terms of SO(8) representations, labeled by their Dynkin weights, with subscripts denoting conformal dimensions.

\subsection{The squashed $S^7$ family in ExFT} \label{s:SquashExFT}
The consistent truncation \eqref{eq:VUV} to the lowest KK multiplet ($\ell=0$ in \eqref{eq:towersS7}) contains the round $S^7$ vacuum (corresponding to $V(x)=\mathbb{I}$) but no other solution of the family of squashed backgrounds \eqref{eq:domain_wall}. Rather, these squashings correspond to excitations of scalars from higher KK levels. Specifically, the squashing \eqref{eq:domain_wall} preserves an ${\rm Sp}(2)\times {\rm Sp}(1)$ subgroup of isometries, which is embedded into the SO(8) isometry group of the round $S^7$ such that the fundamental representations decompose as
\begin{equation} 
\begin{split}
{\rm SO}(8) &\longrightarrow {\rm Sp}(2)\times {\rm Sp}(1) \,,\\
[1,0,0,0] \rightarrow [1,0,1]\,,\quad
[0,0,1,0] &\rightarrow [0,1,0]\oplus[0,0,2]\,,\quad
[0,0,0,1] \rightarrow [1,0,1] \,.
\end{split}
\label{eq:Sp2Sp1}
\end{equation}
A scan of the scalar spectrum \eqref{eq:towersS7} shows that the spectrum contains four scalar fields that are singlet under ${\rm Sp}(2)\times {\rm Sp}(1)$, sitting at KK levels 0, 2, and 4, respectively, descending from the SO(8) representations
\begin{equation}
	\begin{split}
		\ell=0 &: \;\; [0,0,2,0]_{2} \,, \\
		\ell=2 &: \;\; [0,0,0,0]_{6} \oplus [0,2,0,0]_{4} \,, \\
		\ell=4 &: \;\; [2,0,0,2]_{6} \,.
		\label{eq:Sp2Sp1Singlets}
	\end{split}
\end{equation}
The squashed background \eqref{eq:domain_wall} requires a non-vanishing contribution from the $[0,2,0,0]$ at level $\ell=2$~\cite{Duff:1986hr} and, thus, lives beyond the consistent truncation \eqref{eq:VUV}.

However, the embedding of the scalar KK fluctuations according to \eqref{eq:scalarFluc} allows us to explicitly construct different alternative truncations to other subsets of fields which in particular allow to embedd the solution \eqref{eq:domain_wall}. E.g.\ the truncation of the full KK spectrum keeping all singlets under ${\rm Sp}(2)\times {\rm Sp}(1)$ defines a consistent truncation by the standard symmetry argument: by simple representation reasoning, singlet fields can never define non-vanishing sources for non-singlet fields, it is thus consistent to truncate out all the non-singlet fields. This truncation retains one field from the KK tower of gravitino fields, and thus corresponds to a $D=4$, ${\cal N}=1$ theory \cite{Cassani:2011fu}, further discussed in \cite{Nilsson:2023ctq}. In the scalar sector, the truncation to ${\rm Sp}(2)\times {\rm Sp}(1)$ singlets only keeps the four scalar fields \eqref{eq:Sp2Sp1Singlets} and can be described in closed form by integrating the corresponding fluctuations \eqref{eq:scalarFluc} to
\begin{equation} 
{\cal V}(x,y) 
 = \mathring{U}(y)\;{\rm exp}\Big[   \sum_{\rm singlets}  \phi^i(x) \, s_i^{I,\Sigma}\, \cP_{I} \cY_\Sigma(y) \Big]
\,.
\label{eq:TruncationSinglets}
\end{equation}
The index $i$ here labels the four ${\rm Sp}(2)\times {\rm Sp}(1)$ singlets found in the tensor product of the 70 non-compact generators of E$_{7(7)}$ and the scalar harmonics, thereby defining the constant tensor $s_i^{I,\Sigma}$.
The group theoretical structure underlying this truncation can be made more transparent by representing the seven sphere as a coset space~\cite{Bais:1983wc}
\begin{equation} 
S^7 = \frac{{\rm G}}{{\rm H}} = \frac{{\rm Sp}(2)\times {\rm Sp}(1)_0}{{\rm Sp}(1)_{\rm L}\times {\rm Sp}(1)_{\rm D}}\,,
\label{eq:cosetSp2Sp1}
\end{equation}
where the different ${\rm Sp}(1)$ factors are embedded as 
\begin{equation} 
{\rm Sp}(2) \supset {\rm Sp}(1)_{\rm L}\times {\rm Sp}(1)_{\rm R}\;,\qquad
{\rm Sp}(1)_{\rm D}=\big( {\rm Sp}(1)_0 \times {\rm Sp}(1)_{\rm R}\big)_{\rm diag}
\,,
\end{equation}
and for clarity we have added a `$_0$' subscript to the ${\rm Sp}(1)$ subgroup of \eqref{eq:Sp2Sp1}. The seven sphere \eqref{eq:cosetSp2Sp1} can be represented by an $({\rm Sp}(2)\times {\rm Sp}(1))$-valued coset representative $S(y)$, such that the infinitesimal action of the ${\rm Sp}(2)\times {\rm Sp}(1)$ isometry group on the coordinates is realized as
\begin{equation}
\delta_\Lambda S(y) = \Lambda S(y) - S(y) h_\Lambda(y)\;,\qquad
\Lambda \in \mathfrak{sp}(2)\oplus \mathfrak{sp}(1)_0\;,\quad  h_\Lambda(y) \in \mathfrak{sp}(1)_{\rm L}\oplus \mathfrak{sp}(1)_{\rm D}\,.
\label{eq:cosetS}
\end{equation}
The consistent truncation (\ref{eq:TruncationSinglets}) can then be given in more compact form as
\begin{equation} 
{\cal V}(x,y) 
 = \mathring{U}(y)\;S(y)\,W(x)\,S^{-1}(y)
\;,
\label{eq:TruncationCoset}
\end{equation}
with the $({\rm Sp}(2)\times {\rm Sp}(1))$ coset representative $S(y)$ and an E$_{7(7)}$ matrix $W(x)$ defined to live in the commutant of the denominator group ${\rm H}= {\rm Sp}(1)_{\rm L}\times {\rm Sp}(1)_{\rm D}$ in E$_{7(7)}$, i.e.\
\begin{equation} 
W(x)\,h = h\,W(x)
\quad
\forall \,h \in \mathfrak{sp}(1)_{\rm L}\oplus \mathfrak{sp}(1)_{\rm D}
\;.
\label{eq:commutant}
\end{equation}
Indeed, \eqref{eq:cosetS} and \eqref{eq:commutant} imply that the factor $S(y)\,W(x)\,S^{-1}(y)$ in \eqref{eq:TruncationCoset} is invariant under the action of $({\rm Sp}(2)\times {\rm Sp}(1))$ up to a compact gauge transformation acting from the right.
The representation \eqref{eq:TruncationCoset} of the consistent truncation immediately reveals the geometry of its scalar target space, given by the commutant of ${\rm H}= {\rm Sp}(1)_{\rm L}\times {\rm Sp}(1)_{\rm D}$ in E$_{7(7)}/{\rm SU}(8)$ as 
\begin{equation} 
{\cal M}_{\rm scalar}= \frac{{\rm SL}(2)}{{\rm SO}(2)} \times \frac{{\rm SL}(2)}{{\rm SO}(2)}
\,,
\label{eq:SL2SL2}
\end{equation}
a K\"ahler manifold parametrized by the four scalar fields from \eqref{eq:Sp2Sp1Singlets}.\footnote{
The same reasoning shows that the analogous construction based on the coset representation
$S^7 =  {{\rm Sp}(2)}\big/{{\rm Sp}(1)_{\rm L}}$ of the seven sphere, yields an ${\cal N}=4$ consistent truncation to the ${{\rm Sp}(2)}$ singlets with scalar target space given by the commutant of the denominator group ${\rm Sp}(1)_{\rm L}$ in ${\rm E}_{7(7)}/{\rm SU}(8)$:
\begin{equation} 
{\cal M}_{\rm scalar}= \frac{{\rm SL}(2)}{{\rm SO}(2)} \times \frac{{\rm SO}(6,3)}{{\rm SO}(6)\times {\rm SO}(3)}
\,,
\label{eq:SO63}
\end{equation}
as explicitly found in \cite{Cassani:2011fu}.}

The $D=4$ theory can be obtained by plugging the ansatz \eqref{eq:TruncationCoset} into the ExFT Lagrangian of \cite{Hohm:2013uia}. For the bosonic sector, parametrizing the matrix $W$ by four scalar fields $\{u, v, c, \chi\}$, this gives rise to the four-scalar Lagrangian
\begin{equation}\label{eq:4scalar}
	\begin{split}
		|g|^{-1/2}\,{\cal L}_{(4)} &= R_{(4)} -\frac{63}{2}\,\partial_\mu u \,\partial^\mu u -21\,\partial_\mu v \,\partial^\mu v -3\,e^{-6u-2v} \partial_\mu c \,\partial^\mu c -\frac12\,e^{-6u+12v}\,\partial_\mu\chi\,\partial^\mu \chi - V_{\rm pot}\,,\\[1ex]
		V_{\rm pot} &= -6\,e^{-9u+4v}-48\,e^{-9u-3v}+12\,e^{-9u-10v}-72\,e^{-15u-12v}\,c^2 \\
		& \quad -12\,e^{-15u+2v}\,(c+\chi)^2-18\,e^{-21u}\,\big(1+c^2+2\,c\,\chi\big)^2 \,.
	\end{split}
\end{equation}
It is straightforward to verify that the intersection of the scalar target space \eqref{eq:SL2SL2} with the ${\rm SL}(8)$ of \eqref{eq:Freund-Rubin} capturing the Freund-Rubin solutions corresponds to a further consistent truncation to two scalar fields which precisely reproduces the result \eqref{eq:potential} of \cite{Page:1983mke}.
The ${\cal N}=1$ consistent truncation to four scalars \eqref{eq:4scalar} has already been found in \cite{Cassani:2011fu}. What is new in the present construction is its explicit embedding \eqref{eq:TruncationCoset} via a twist matrix in ExFT which allows us to apply the universal mass formulas such as \eqref{eq:MassMatrixScalarClean}, \eqref{eq:MNM} to any background within this truncation. With the frame given by
\begin{equation} 
{\cal V} = \mathring{U}\,S\,W\,S^{-1}
\,,
\label{eq:frame}
\end{equation}
a general background will satisfy \eqref{eq:GenLeib}, but, in general, with $y$-dependent intrinsic torsion $X_{\fl{AB}}{}^{\fl{C}}(y)$.\footnote{
Because of the scalar fields in $W(x)$, the intrinsic torsion will also depend on $x$, but this is standard already in gauged supergravity. In particular, it does not interfere with the generalised (Leibniz) parallelisability. } Thus, it is no longer generalised Leibniz parallelisable, but still generalised parallelisable. As a consequence the mass formulae \eqref{eq:MassMatrixScalarClean}, \eqref{eq:MNM} still apply. This has been used in \cite{Duboeuf:2022mam} in order to derive the full Kaluza-Klein spectrum around the ${\cal N}=1$, AdS$_4\times S_{\rm squashed}^7$ vacuum. In the following, we will extend this analysis to also derive quadratic couplings around the domain wall background \eqref{eq:domain_wall}, or any other background of the consistent truncation.

\section{Couplings around the domain wall background} \label{s:KK}
As we reviewed in section \ref{s:GLPandMass}, we can efficiently describe the linearised fluctuations around the round $S^7$ using its generalised Leibniz parallelisation in ExFT. Similarly, using the generalised parallelisation of the family of squashed $S^7$'s of section \ref{s:SquashExFT}, we obtain a simple expression of the linearised fluctuations around the family of squashed $S^7$ that describe the flow of section \ref{s:SquashedS7}. While this was used in \cite{Duboeuf:2022mam} to obtain the full Kaluza-Klein spectrum of the squashed $S^7$, here we will further extend the computation to obtain the quadratic couplings of the higher Kaluza-Klein modes around the domain wall solution. These couplings encode all the information about the holographic 2-point functions along the dual RG flow.

In computing the quadratic couplings, we have to choose a field basis for the Kaluza-Klein fluctuations, or choice of ``frame''. Two natural choices are to multiply the generalised vielbein of the 2-scalar truncation, $\sV(x,y)$ of \eqref{eq:TruncationCoset}, from the right or left, leading to different kinetic terms for the scalars $u$, $v$ and the Kaluza-Klein fluctuations. The kinetic term of ExFT is given by$D_{\mu}M^{MN} D^\mu M_{MN}$ with $M=\sV \sV^T$, which can be expressed in terms of the current $J_{\mu}(\sV)_\uM {}^{\uN}=(\sV^{-1}D_\mu \sV)_\uM{}^\uN$ as
\begin{equation} \label{eq:kinterm}
	D_{\mu}M^{MN} D^\mu M_{MN} = - 2\left(\Tr \,J^\mu J_\mu ^T +\Tr\,J^\mu J_\mu \right) \,.
\end{equation}

Let us denote the Kaluza-Klein fluctuations by the matrix $\delta= \exp[ \cP_{I} \sum_\Sigma  j^{I,\Sigma}(x) \,  \cY_\Sigma(y)]$. Then, it is easy to see that $J_\mu$ becomes
\begin{equation}\label{eq:currents}
	\begin{split}
		J_\mu(\sV\delta )&=\delta^{-1}J_\mu(\sV)\delta+J_\mu(\delta) \,, \\
		J_\mu(\delta\sV)&=\sV^{-1}J_\mu(\delta)\sV+J_\mu(\sV) \,,
	\end{split}
\end{equation}
depending on whether we multiply the generalised $\sV$ with the Kaluza-Klein fluctuations from the left or the right. We clearly see that multiplying $\sV$ from the left leaves the kinetic term for the background, i.e.\ the kinetic term of the scalars $u$, $v$, invariant, but introduces $u$, $v$ factors into the kinetic term of the Kaluza-Klein fluctuations. On the other hand, using $\sV\, \delta$, we find that the Kaluza-Klein fluctuations enter the kinetic terms of $u$, $v$ and hence introduce corrections to the kinetic terms of \eqref{eq:4scalar}. In order to simply retain the normalisation \eqref{eq:4scalar}, we will choose to define our Kaluza-Klein fluctuations as multiplying the truncation $\sV$ from the left. Note that this differs from the conventions of the Kaluza-Klein spectrometry \cite{Malek:2019eaz,Malek:2020yue,Duboeuf:2022mam}, used in \eqref{eq:scalarFluc}.
The quadratic couplings around the domain wall solutions are then straightforwardly obtained by evaluating the action of the mass operator (\ref{eq:MassMatrixScalarClean}) in this basis.

Before presenting our results for the quadratic couplings, let us recall some facts about the spectrum on the left-squashed $S^7$ from \cite{Nilsson:2018lof,Ekhammar:2021gsg,Karlsson:2021oxd,Duboeuf:2022mam,Karlsson:2023dnl}. The Kaluza-Klein spectrum is organized into long ${\cal N}=1$ supermultiplets $L[J,\Delta]$ in various $\Sp{2}\times \Sp{1}$ representations. These multiplets are defined by the spacetime spin $J$ and conformal dimension $\Delta$ of their superconformal primaries, given by the universal formula \cite{Duboeuf:2022mam}
\begin{equation} \label{eq:SquashedSpectrum}
	\Delta_{J,s} = 1 +\frac53 s+\frac 13\sqrt{(3J+2s^2)^2+5\,{\cal C}_3} \,.
\end{equation}
Here ${\cal C}_3$ is a combination of the $\Sp{2}$ and $\Sp{1}$ quadratic Casimirs, which, for a generic representation with Dynkin labels $[p,q,r]$, is given by
\begin{equation}
	\begin{split}
		{\cal C}_3&={\cal C}(p,q)+3{\cal C}(r)\\
		&=\frac12(p^2+2q^2+4p+6q+2pq)+\frac34r(r+2) \,.
	\end{split}
\end{equation}	
The extra parameter $s$ in \eqref{eq:SquashedSpectrum} is a half integer which, for most Kaluza-Klein towers, surprisingly behaves like an $\SL{2}$ spin. This motivates the following notational convention of \cite{Duboeuf:2022mam} 
\begin{equation}
\begin{split}
L[J]\smallotimes[s]&=\bigoplus_{i=0}^{2s} L[J,\Delta_{J,i-s}] \,, \\
L[J]\smallotimes\{s_1,\cdots,s_n\} &= \bigoplus_{i=1}^{n}L[J,\Delta_{J,s_i}] \,.
\end{split}
\end{equation}

The Kaluza-Klein spectrum organises itself into towers of the form
\begin{equation}\label{eq:spectrumreps}
\begin{split}
[k,q,k]_{k>1,q>1}\,&:\quad L[\tfrac32] \smallotimes [0]\oplus L[1] \smallotimes [\tfrac12]\oplus L[\tfrac12] \smallotimes [\tfrac12\smallotimes \tfrac12]\oplus L[0] \smallotimes [\tfrac12 \smallotimes 1] \,, \\
[k,q,k+2]_{k>0,q>1}\,\&\, [k+2,q,k]_{k>0,q>0}\,&:\quad L[0] \smallotimes[\tfrac12]\oplus L[1] \smallotimes [\tfrac12]\oplus L[\tfrac12] \smallotimes [\tfrac12\smallotimes \tfrac12] \,,\\
[k,q,k+4]_{q>1}\,\&\,[k+4,q,k]\,&:\quad L[\tfrac12]\smallotimes[0]\oplus L[0]\smallotimes[\tfrac12] \,,
\end{split}
\end{equation}
which degenerate for some small values of representations, such as
\begin{equation} \label{eq:012rep}
[0,1,2]\,:\quad L[1] \smallotimes [\tfrac12]\oplus L[\tfrac12] \smallotimes \{-1,+1\}\,.
\end{equation}
For more details and to see the remaining multiplets consult \cite{Duboeuf:2022mam}.

Since the consistent truncation (and thus domain wall) preserves $\Sp{2} \times \Sp{1}$, different $\Sp{2} \times \Sp{1}$ representations will not mix along the flow. Therefore, we can restrict our attention to any fixed $\Sp{2} \times \Sp{1}$ representation to compute their quadratic couplings. 
As an illustration of our method, we will evaluate the mass operator (\ref{eq:MassMatrixScalarClean}) on the scalars fields 
 in the $[0,0,0]$, the $[0,1,2]$ and $[0,2,4]$ representations, which at the squashed $S^7$ sit in the long multiplets \eqref{eq:012rep} and \eqref{eq:spectrumreps} as follows 
\begin{equation}\label{eq:multiplets}
	\begin{split}
		L[\tfrac12,\Delta]: \lambda{\phantom{_\mu}} &\xrightarrow{~Q~} \,A_\mu\,  \oplus  \phi\phantom{_\mu} \xrightarrow{~Q~} \lambda \,, \\
		L[0,\Delta]: \phi{\phantom{_\mu}} &\xrightarrow{~Q~} \phantom{A_\mu}\;\; \lambda \phantom{+\,}\;\; \xrightarrow{~Q~} \phi \,.
	\end{split}
\end{equation}

\subsection{$[0,0,0]$ sector}

As a warm up, let us compute the quadratic couplings in the sector of scalar fields singlet under $\Sp{2} \times \Sp{1}$, forming the bosonic sector of the ${\cal N}=1$ truncation of 
\cite{Cassani:2011fu,Nilsson:2023ctq}. This sector carries four scalar field fluctuations whose basis we label as $\varphi_i$. 
The computation is straightforward: we first compute the intrinsic torsion from (\ref{eq:frame}), where $W$ is evaluated on the domain wall solution. Next we build the 
associated mass operator (\ref{eq:MassMatrixScalarClean}) and evaluate its action on the scalar fluctuations in this sector.
The final result for the quadratic fluctuations in this sector takes the form
	\begin{align}\label{eq:000sec}
		{\cal L}_{[0,0,0]} &=
		-\frac{1}{2}\, \left(\partial_\mu \varphi_1 \partial^\mu \varphi_1 + \partial_\mu \varphi_2 \partial^\mu \varphi_2\right)
		-\frac{1}{2} \,e^{-6 u-2 v}\,\partial_\mu \varphi_3 \partial^\mu \varphi_3
		-\frac{1}{2} \,e^{-6 u+ 12 v} \,\partial_\mu \varphi_4 \partial^\mu \varphi_4 \notag
		\\
		& \quad
		-\left(54\,e^{-21 u} +16 \, e^{-9 u-10 v}-36\,  e^{-9 u-3 v}-2\, e^{-9 u+4 v}\right)\varphi_1^2
 -12\,e^{-15 u+2 v}	\,  \varphi_4^2
  \notag\\
   &\quad
		+2 \sqrt{6}\left(9\, e^{-21 u}-4\,  e^{-9 u-10 v}
		 -  e^{-9 u+4 v}\right) \varphi_1 \varphi_2
 -3 \left(3 \, e^{-21 u} +2 \, e^{-9 u-10 v}-e^{-9 u+4 v}\right)\varphi_2^2
      \notag\\
   &\quad
   -2\left( 3 \, e^{-21 u}+6 \, e^{-15 u-12 v}+ e^{-15 u+2 v}\right) \varphi_3^2
-4\sqrt{6} \left(  3 \, e^{-21 u}+e^{2 v-15 u}\right) \varphi_3 \varphi_4 
 	\,,
\end{align}
where $u$ and $v$ denote the domain wall solution constructed in section~\ref{s:SquashedS7} above.
One verifies that at the endpoints of the flow, the Lagrangian (\ref{eq:000sec}) reproduces the masses of the scalar fields for the squashed and the round $S^7$.
In fact, since all four scalar singlets lie within the ${\cal N}=1$ truncation (\ref{eq:SL2SL2}), we could have arrived at this result directly by linearising the potential (\ref{eq:4scalar}).
It is a good consistency check of our method, that the results indeed agree. 
In contrast, the scalar fluctuations at higher KK levels do not lie within a consistent truncation and the respective couplings can only 
be obtained by the mass operator (\ref{eq:MassMatrixScalarClean}), as we shall discuss in the following.
We finally note that in order to study the complete set of fluctuation equations in this singlet sector, 
one also has to take into account the fluctuations of the metric around the AdS$_4$ background.

\subsection{$[0,1,2]$ sector}
Let us now extend the computation to scalar modes that do not lie within the consistent truncation. As an example, we choose the sector in the
$[0,1,2]$ representation of $\Sp{2} \times \Sp{1}$. This is the representation of the vector fields that become massless on the round $S^7$. As a consequence, 
the associated scalar Goldstone modes on the squashed sphere become physical scalars on the round sphere. 
There are in total eight scalar fluctuations $\phi _i$ transforming in the  $[0,1,2]$ representation. 
Repeating the above computation in this sector leads to the Lagrangian
\begin{align}\label{eq:012sec}
		{\cal L}_{[0,1,2]} &=
		-\frac{1}{2}\, e^{-7 v} \left(\partial_\mu \phi_1 \partial^\mu \phi_1 + \partial_\mu \phi_2 \partial^\mu \phi_2\right)
		-\frac{1}{2} \,e^{-6 u-9 v}\,\partial_\mu \phi_3 \partial^\mu \phi_3
		-\frac{1}{2} \,e^{-6 u-2 v} \,\partial_\mu \phi_4 \partial^\mu \phi_4 \notag
		\\
		& \quad
		-\frac{1}{2} \, e^{-6 u+5v} \left( \partial_\mu \phi_5 \partial^\mu \phi_5+\partial_\mu \phi_6 \partial^\mu \phi_6 \right)
		-\frac{1}{2} \,e^{-12u-4 v}\,\partial_\mu \phi_7 \partial^\mu \phi_7
		-\frac{1}{2} \,e^{-12 u+3 v} \,\partial_\mu \phi_8 \partial^\mu \phi_8 \notag \\
		& \quad 
		-\left(10\,e^{-9 u-10 v}-2\,e^{-9 u-3 v}\right) \phi _1^2 
		-4 \,e^{-15 u-5 v}\, \phi_{345}^2 
		-4 \sqrt{2} \left( e^{-15 u-5 v}+3e^{-21 u} \right) \phi _6  \phi_{345} \notag \\
		& \quad
		-\left(
		6 \,e^{-15 u-5 v}+4\,e^{-15u+2 v}+4 \,e^{-15 u+9 v}+6 \,e^{-21 u} \right)
		\phi _6^2
		\,,
\end{align}
with $\phi_{345}=\phi_3+\phi_4+\phi_5$. The result shows that the potential only depends on three out of the eight $\phi_i$'s. Therefore, the remaining five fluctuations are Goldstone modes along the flow. It is interesting to compare this with the squashed sphere endpoint. From equations \eqref{eq:spectrumreps} and \eqref{eq:multiplets} we see that at the squashed $S^7$ vacuum there are only two physical scalars in the $[0,1,2]$ in the spectrum. Indeed, plugging in the values of $u$ and $v$ for the squashed $S^7$ \eqref{eq:S7S7}, we see that $\phi_1$ additionally drops out of the potential at the squashed $S^7$ endpoint, thus reproducing the expected number of Goldstone scalars. Moreover, the potential \eqref{eq:012sec} reproduces the masses of the scalar fields at the squashed and round $S^7$ points.

\subsection{$[0,2,4]$ sector}
As a last example, we study the six scalars $\psi _i$ in the $[0,2,4]$ representation. Their quadratic couplings are obtained 
by the analogous computation and read
\begin{align}\label{eq:024sec}
{\cal L}_{[0,2,4]} &=
-\frac{1}{2}\,e^{-7 v}\,\partial_\mu \psi_1 \partial^\mu \psi_1
-\frac{1}{2} \,e^{-6 u-2 v}  \left(\partial_\mu \psi_3 \partial^\mu \psi_3 + \partial_\mu \psi_4 \partial^\mu \psi_4\right)
-\frac{1}{2}\left( \partial_\mu \psi_5 \partial^\mu \psi_5+\partial_\mu \psi_6 \partial^\mu \psi_6 \right)
\notag\\
&\quad
-\frac{1}{2} \,e^{-6u+5 v}\,\,\partial_\mu \psi_2 \partial^\mu \psi_2
-6\,e^{-9u} \left(3\,  e^{-10 v}
+ e^{-3 v}\right) \psi _1^2
-8 \,e^{-9 u-10 v}\left(
\sqrt{5} \,\psi _5 
- \sqrt{3} \, \psi _6 \right)  \psi _1
\notag\\
&\quad
-4\, e^{-9 u} \left(
e^{-10 v}
+5 \,e^{-3 v}
+ e^{4 v}  \right) \psi _5^2 
-4\, e^{-9 u} \left(
e^{-10 v}
- e^{-3 v}
+3 \, e^{4 v} \right) \psi _6^2
\notag\\
&\quad
-\left( 6 \, e^{-15 u-5 v}
+8\, e^{-15u+2 v}
+12 \,e^{-21 u} \right) \tilde\psi_3^2
-
\left(
10\, e^{-15 u-5 v}
+8\, e^{-15u+2 v}
-12 \,e^{-21 u} \right)
\tilde\psi_4^2
\notag\\
& \quad
+4 \sqrt{15} \,e^{-15 u-5 v}\,\tilde\psi_3 \tilde\psi_4
\,,
\end{align}
where
$\tilde\psi_3=\psi_3-\frac12\sqrt{3}\,\psi_2$,
$\tilde\psi_4=\psi_4 -\frac12\sqrt{5}\,\psi_2$\,.
The potential only depends on five out of the six scalars. Therefore, one scalar field will be a Goldstone mode
along the entire flow. 
Finally, as a consistency check, the potential \eqref{eq:024sec} again reproduces the masses at both of the endpoints.

\subsection{Comments}

We have shown how the mass operator (\ref{eq:MassMatrixScalarClean}) allows to determine the quadratic couplings of scalar fluctuations
around the domain wall background constructed in section~\ref{s:SquashedS7}. We have restricted to spelling out three examples, but the method of course 
extends to all higher KK levels.
The resulting couplings (\ref{eq:000sec}), (\ref{eq:012sec}), (\ref{eq:024sec}), carry the full information for the computation of the holographic 2-point functions 
along the dual RG flow. The corresponding computation requires a careful setup of the holographic renormalisation procedure along the lines of~\cite{Bianchi:2001de,Bianchi:2001kw,Papadimitriou:2004rz}, and will be interesting to take on in the future.

\section{Conclusions} \label{s:Conclusions}
In this paper, we revisited the problem of constructing a domain wall solution interpolating between the ${\cal N}=1$ AdS$_4 \times S^7_{\rm squashed}$ and ${\cal N}=8$ AdS$_4 \times S^7_{\rm round}$ vacua of 11-dimensional supergravity. There is no supersymmetric domain wall preserving the $\Sp{2} \times \Sp{1} \subset \SO{8}$ isometry of the squashed $S^7$, as can, for example, be seen by noting that the ${\cal N}=8$ gravitini of the round $S^7$ are not $\Sp{2} \times \Sp{1}$ singlets. Instead, we construct an explicit non-supersymmetric flow by using a consistent truncation to 4-dimensional ${\cal N}=1$ supergravity \cite{Page:1984ad,Cassani:2011fu} and solving the second-order flow equations numerically. Interestingly, within this ${\cal N}=1$ truncation, the ${\cal N}=8$ round $S^7$ appears non-supersymmetric, since its massless gravitini reside amongst the higher KK modes.

Extending recent techniques from ExFT \cite{Malek:2019eaz,Malek:2020yue,Duboeuf:2022mam}, we were able to compute the quadratic couplings of KK fluctuations around the domain-wall solution (or any other solution of the 2-dimensional truncation). This relied on the fact that the family of squashed $S^7$ still admits a trivialisation of the generalized tangent bundle \cite{Duboeuf:2022mam}, allowing us to construct a globally well-defined generalised frame ${\cal V}(x,y)$. However, the intrinsic torsion $X_{\uM \uN}{}^\uP(y)$ of this generalised frame is not constant, reflecting the fact that the squashed $S^7$ only admits consistent truncations to ${\cal N}=4$ or ${\cal N}=1$ but not ${\cal N}=8$. We presented the quadratic couplings of some low-lying KK scalars, but the same method can be applied to any other KK tower. This encodes the information needed to extract all the holographic two-point functions along the flow.

Our work opens up several natural directions for future investigations. One would be to understand the analytic structure underlying the quadratic couplings of section \ref{s:KK}, similar to the group-theoretic formula that appear in the KK spectra of vacua with sufficient isometries, e.g. \cite{Malek:2020yue,Guarino:2020flh}. For this it would be useful to extend our computation of the quadratic couplings to other KK towers. Another interesting question would be to push the calculations of section \ref{s:KK} to cubic order, giving access to 3-point functions along the RG flow on the field theory side. These results yield the relevant input for the holographic renormalisation procedure~\cite{Bianchi:2001de,Bianchi:2001kw,Papadimitriou:2004rz}, that would allow us to precisely obtain the field theory correlators along the flow.

Finally, our results here, together with \cite{Duboeuf:2022mam}, provide the first extension of the ExFT formalism of Kaluza-Klein spectroscopy \cite{Malek:2019eaz,Malek:2020yue} beyond generalised parallelisable spaces, i.e.\ vacua of maximally supersymmetric consistent truncations. While the family of squashed $S^7$'s still admits a generalised parallelisation, there is no longer an ${\cal N}=8$ consistent truncation that the solutions belong to. Hence our method here provides a first window into how to generalise the Kaluza-Klein spectroscopy to more general truncations that break some supersymmetry.

All of these questions will allow us to address the ultimate question: how does having a consistent truncation affect the structure of correlation functions in the holographic dual? We leave these exciting research endeavours for the future. 

\section*{Acknowledgements}
We are grateful to Nikolay Bobev, Davide Cassani, Jerome Gauntlett, Joel Karlsson, Gabriel Larios, Michela Petrini, and Dan Waldram for useful discussions and correspondence. MG is grateful for the hospitality of ENS Lyon. MG and EM are supported by the Deutsche Forschungsgemeinschaft (DFG, German Research Foundation) via the Emmy Noether program ``Exploring the landscape of string theory flux vacua using exceptional field theory'' (project number 426510644).


\providecommand{\href}[2]{#2}\begingroup\raggedright\endgroup

\end{document}